\documentclass[twocolumn,tighten]{aastex63}
\usepackage{amsmath,mathtools,mathrsfs}
\usepackage{hyperref}
\usepackage{verbatim}
\usepackage{xspace}
\usepackage{longtable}
\usepackage{multirow}
\usepackage{afterpage}
\usepackage{booktabs}

\def\muas{\mu{\rm as}} 
\def\dd{\mathrm{d}}
\def\m87{M~87*}

\newcommand{\VIDA}{\texttt{VIDA}\xspace}
\newcommand{\ReX}{\texttt{REx}\xspace}
\newcommand{\ehtim}{\texttt{eht-imaging}\xspace}

\newcommand{\C}{\texttt{C}\xspace}

\newcommand{\msun}{{\rm M_{\odot}}}

\renewcommand{\L}{{\mathcal L}}

\defcitealias{EHTCI}{Paper~I}
\defcitealias{EHTCII}{Paper~II}
\defcitealias{EHTCIII}{Paper~III}
\defcitealias{EHTCIV}{Paper~IV}
\defcitealias{EHTCV}{Paper~V}
\defcitealias{EHTCVI}{Paper~VI}

\shorttitle{The Ellipticty of \m87}
\shortauthors{Tiede et al.}

\begin{document}
  
\title{Measuring the Ellipticity of \m87 Images}


\correspondingauthor{Paul Tiede}
\email{ptiede@perimeterinstitute.ca}

\author[0000-0003-3826-5648]{Paul Tiede}
\affiliation{ Black Hole Initiative at Harvard University, 20 Garden Street, Cambridge, MA 02138, USA}
\affiliation{ Center for Astrophysics | Harvard \& Smithsonian, 60 Garden Street, Cambridge, MA 02138, USA}
\affiliation{ Perimeter Institute for Theoretical Physics, 31 Caroline Street North, Waterloo, ON, N2L 2Y5, Canada}
\affiliation{ Department of Physics and Astronomy, University of Waterloo, 200 University Avenue West, Waterloo, ON, N2L 3G1, Canada}
\affiliation{ Waterloo Centre for Astrophysics, University of Waterloo, Waterloo, ON N2L 3G1 Canada}

\author[0000-0002-3351-760X]{Avery E. Broderick}
\affiliation{ Perimeter Institute for Theoretical Physics, 31 Caroline Street North, Waterloo, ON, N2L 2Y5, Canada}
\affiliation{ Department of Physics and Astronomy, University of Waterloo, 200 University Avenue West, Waterloo, ON, N2L 3G1, Canada}
\affiliation{ Waterloo Centre for Astrophysics, University of Waterloo, Waterloo, ON N2L 3G1 Canada}

\author[0000-0002-7179-3816]{Daniel C. M. Palumbo}
\affiliation{ Black Hole Initiative at Harvard University, 20 Garden Street, Cambridge, MA 02138, USA}
\affiliation{ Center for Astrophysics | Harvard \& Smithsonian, 60 Garden Street, Cambridge, MA 02138, USA}

\author[0000-0003-2966-6220]{Andrew Chael}
\altaffiliation{NASA Hubble Fellowship Program Einstein Fellow}
\affiliation{Princeton Center for Theoretical Science, Princeton University, Jadwin Hall, Princeton, NJ 08544, USA}

\begin{abstract}
    The Event Horizon Telescope (EHT) images of the supermassive black hole at the center of the galaxy M 87 provided the first image of the accretion environment on horizon scales. General relativity predicts that the image of the shadow should be nearly circular given the inclination angle of the black hole M 87*. A robust detection of ellipticity in the image reconstructions of M 87* could signal new gravitational physics on horizon scales. Here we analyze whether the imaging parameters used in EHT analyses are sensitive to ring ellipticity and measure the constraints on the ellipticity of M 87*. We find that the top set is unable to recover ellipticity. Even for simple geometric models, the true ellipticity is biased low, preferring circular rings. Therefore, to place a constraint on the ellipticity of M 87*, we measure the ellipticity of 550 simulated data sets of GRMHD simulations. We find that images with intrinsic axis ratios of 2:1 are consistent with the ellipticity seen from the EHT image reconstructions.
\end{abstract}


\section{Introduction}
\label{sec:intro}

The Event Horizon Telescope (EHT) can resolve the emission around the event horizon of the supermassive black hole \m87, and directly measured its mass for the first time in 2019 \citep[][hereafter \citetalias{EHTCI, EHTCII, EHTCIII, EHTCIV, EHTCV, EHTCVI} respectively]{EHTCI, EHTCII, EHTCIII, EHTCIV, EHTCV, EHTCVI}. Using both imaging and modeling techniques, a ring-like emission structure was observed. The measured ring radius is consistent with a central black hole with mass $6.5\times 10^9\msun$. The direct mass measurement was the first direct observation of the accretion disk on horizon scales. In addition to the size of ring, which correlates to the mass \citep{EHTCVI}, the shape or ellipticity of the ring is theoretically interesting. 

There are a number of factors that could cause ellipticity in the measured ring. First are the shadow \citep{Falcke2000:shadow} and $n=1$ photon ring\footnote{We follow \citep{universalsig} definition where $n=1$ corresponds to photon that have done a 1/2-orbit around the black hole.} \citep{universalsig, Gralla2020}, both of which are related to the existence of spherical photons orbits around Kerr black holes. Given the low inclination of \m87 \citep{Mertens2016}, general relativity (GR) predicts that the observed shadow and photon ring should be highly symmetric. Due to this symmetry, it has been suggested to use image ellipticity to constrain deviations from general relativity near the event horizon. If the no-hair theorem breaks down near the event horizon of supermassive black holes, the shadow may appear more elliptic \citep[see, e.g.,][]{Johannsen2010, Broderick2014NoHair, Johannsen2016TestGR, medeiros2020parametric}.

A second source of non-circularity could come from the imprint of the horizon on the image and has been called the \emph{inner shadow} \citep{Dokuchaev, ChaelInnerShadow}. The inner shadow occurs from light rays that do not pass through the equatorial plane before hitting the horizon. Generally, the emission from the black hole must be concentrated in the equatorial plane for the inner shadow to be visible. For instance, MAD general relativistic magneto-hydrodynamic simulations (GRMHD), which are preferred for \m87 \citep{EHTCM87VIII}, tend to display the inner shadow feature \citep{ChaelInnerShadow}. Interestingly, the location of the inner shadow relative to the photon ring is a function of spin and inclination. During image reconstructions, the displacement of the inner shadow could manifest as a source of non-circularity in the image reconstruction.

A third, more mundane origin of the non-circularity comes from the accretion flow itself. The accretion flow is expected to be highly turbulent and can cause extended features in the image, creating highly elliptical ring reconstructions. Therefore, any measurement of ring ellipticity needs to account for the accretion through calibration or other means.

To constrain the ellipticity of \m87, there are two forms of uncertainty that need to be considered: astrophysical/accretion noise and image uncertainty. While the EHT can resolve the horizon scale structure, its dynamic range and visibility coverage is poor \citepalias{EHTCII, EHTCIII}. Therefore, there are infinitely many images that can reproduce observations. This uncertainty makes measuring the ellipticity of the ring uncertain and requires measuring an ensemble of images to quantify how well we can constrain ellipticity. In \citetalias{EHTCIV}, regularized maximum likelihood (RML) methods were applied to the \m87 data. RML introduces additional assumptions through regularizers that enforce features such as image smoothness \citep{Bouman_2016,Chael_2016,Kuramochi_2018}, sparseness \citep{Wiaux2009a,Wiaux2009b,Honma2014,Akiyama_2017b}, and similarity to some fiducial image \citep{NarayanMEM}. However, how best to choose the weights and functional forms of these regularizers is unknown. To combat this problem, \citetalias{EHTCIV} performed a parameter survey using simulated data sets similar in quality to the 2017 \m87 data. The parameter surveys were then used to assess the relative performance of the different regularizer weights and cuts were placed on different choices based on reconstructed image fidelity metrics to give the optimal ``top set.'' The final set of images are considered the set on which feature extraction and ring ellipticity should be measured. We note that while these images do give a distribution of images, they do not form a posterior. As a result, interpreting quantitative measurements often require some apriori calibration to assess the reliability.

In \citetalias{EHTCVI}, a preliminary attempt at measuring the ellipticity of the image reconstructions of \m87 was presented in Figure 18. To measure the ellipticity, they measured the fractional spread in the ring radius from the reconstructions and found it was similar to the intrinsic images having an axis ratio of $4:3$. This axis ratio suggested that the \m87 images were highly symmetric. However, there are two potential issues. First, the parameter survey used in \citetalias{EHTCIV} did not include an elliptic ring in simulated tests. Therefore, the reliability of the ellipticity measured in the images is unclear. Second, the results presented in \citetalias[Figure 18. of][]{EHTCVI} did not measure the ellipticity of the reconstructed GRMHD images, but rather the actual true image blurred to $20\muas$. Whether the imaging parameter survey used in \citetalias{EHTCIV} can reliably recover ellipticity and how to interpret the measured ellipticity of \m87 are then open questions.

In this paper, we assess whether the \m87 top set used in \citetalias{EHTCIV} can constrain the ellipticity of the \m87 shadow across a set of simulated data tests. We consider a new set of geometric tests specifically targeted to measure ellipticity to evaluate whether the surveys can reliably recover image ellipticity. We then use the image feature extraction tools \ReX \citep{ChaelThesis} and \VIDA \citep{vida} to measure the ellipticity of the reconstructions. In the first part of the paper we show that the measured ellipticity of image reconstructions do not reliably recover the true on-sky value. To overcome these issues, we then run the imaging surveys on a set of GRMHD simulations to calibrate the imaging ellipticity bias. This is similar in spirit to mass calibration procedure done in \citetalias{EHTCVI}, and is necessary to interpret the ellipticity results from the imaging pipelines.

The layout of the paper is as follows:
In \autoref{section:imagefeat} we review the \m87 imaging top set, feature extraction techniques and \m87 results. In \autoref{section:asym_cal_geom} we explore whether the current parameter survey used in \m87 is able to recover an elliptical geometric ring. Afterward, we use a set of GRMHD simulations to calibrate the image reconstruction ellipticity of the \m87 results. In \autoref{sec:conclusions} we review the results and provide an ellipticity constraint for \m87.

\section{Background}\label{section:imagefeat}
This section will review the standard image reconstruction techniques used by the EHT in \citetalias{EHTCIV}. The imaging techniques used in this paper will be identical to the \ehtim pipeline used in \citetalias{EHTCIV}. We will then review the two feature extraction techniques used in this paper, \ReX and \VIDA. These feature extraction techniques are needed since imaging is non-parametric. Therefore, an additional processing step is needed to extract image features of interest, e.g., ring ellipticity. Finally, we will apply \ReX and \VIDA to the \m87 top set. We will reproduce the results from \citetalias{EHTCVI} and extend the analysis to include the orientation of \m87's ellipticity.

\begin{figure}[!t]
    \centering
    \includegraphics[width=\linewidth]{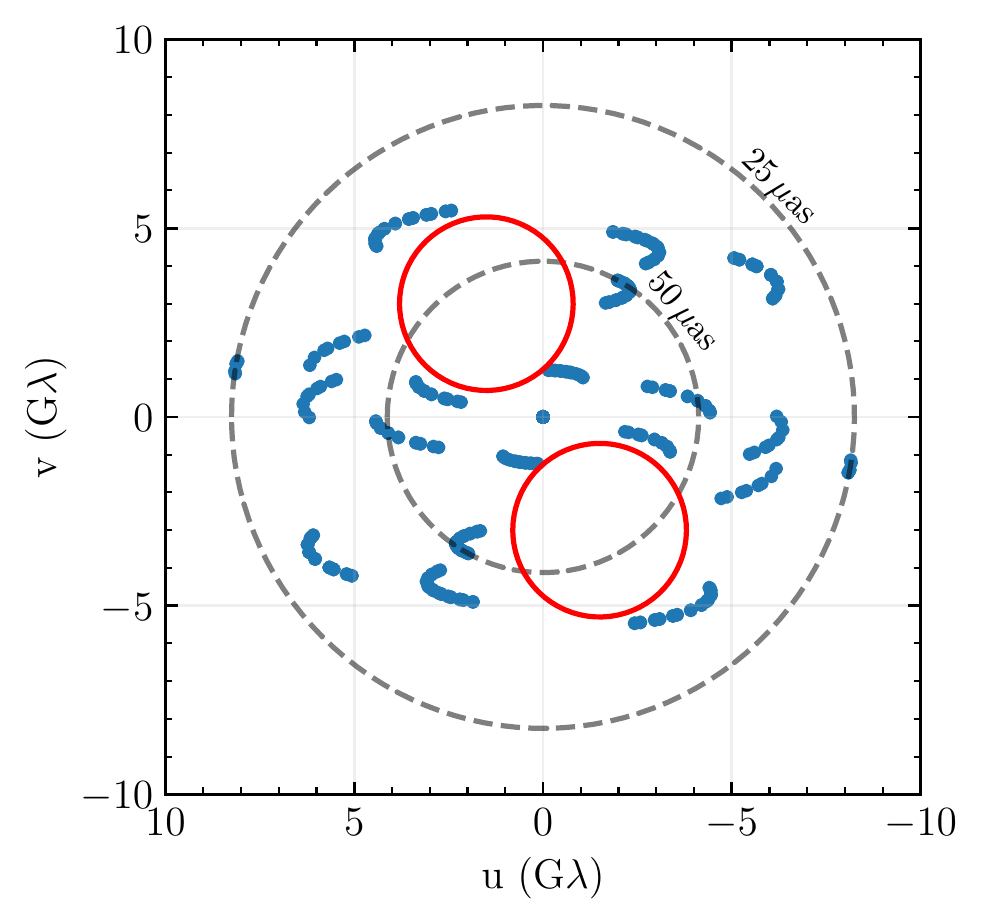}
    \caption{UV coverage of the EHT 2017 array on April 11. The blue dots show where the 2017 array samples in the uv plane in units of G$\lambda$. The black dotted lines show the characteristics location of the image features of $50\,\muas$ and $25\,\muas$ in the uv domain. The red circles highlight the coverage gap in \m87 in the north-south direction.}
    \label{fig:coverage}
\end{figure}
\subsection{Image Reconstructions and the \m87 top set}\label{ssection:imaging}
In this paper, we will focus on the regularized maximum likelihood methods used in \citetalias{EHTCIV}. These methods attempt to make imaging tractable by forward modeling the image, $I$, and minimizing the objective function:
 \begin{equation}\label{eq:img-objective}
     J(I) = \sum_{\rm data} \alpha_d \chi^2_d(I) - \sum_{\rm regularizers}\beta_r S_r(I).
 \end{equation}
Following \citetalias{EHTCIV}, each $\chi^2$ is defined solely from the data products from the EHT telescope, e.g., complex visibilities. The second term encapsulates the additional assumptions or regularizers placed on the image. The $\alpha_d\, \beta_r$ are the ``hyperparameters'' that control the relative weighting of the regularizers and data products. For a list of the regularizers used, see \citetalias{EHTCIV}.
 
\begin{figure*}[!t]
    \centering
    \includegraphics[width=\textwidth]{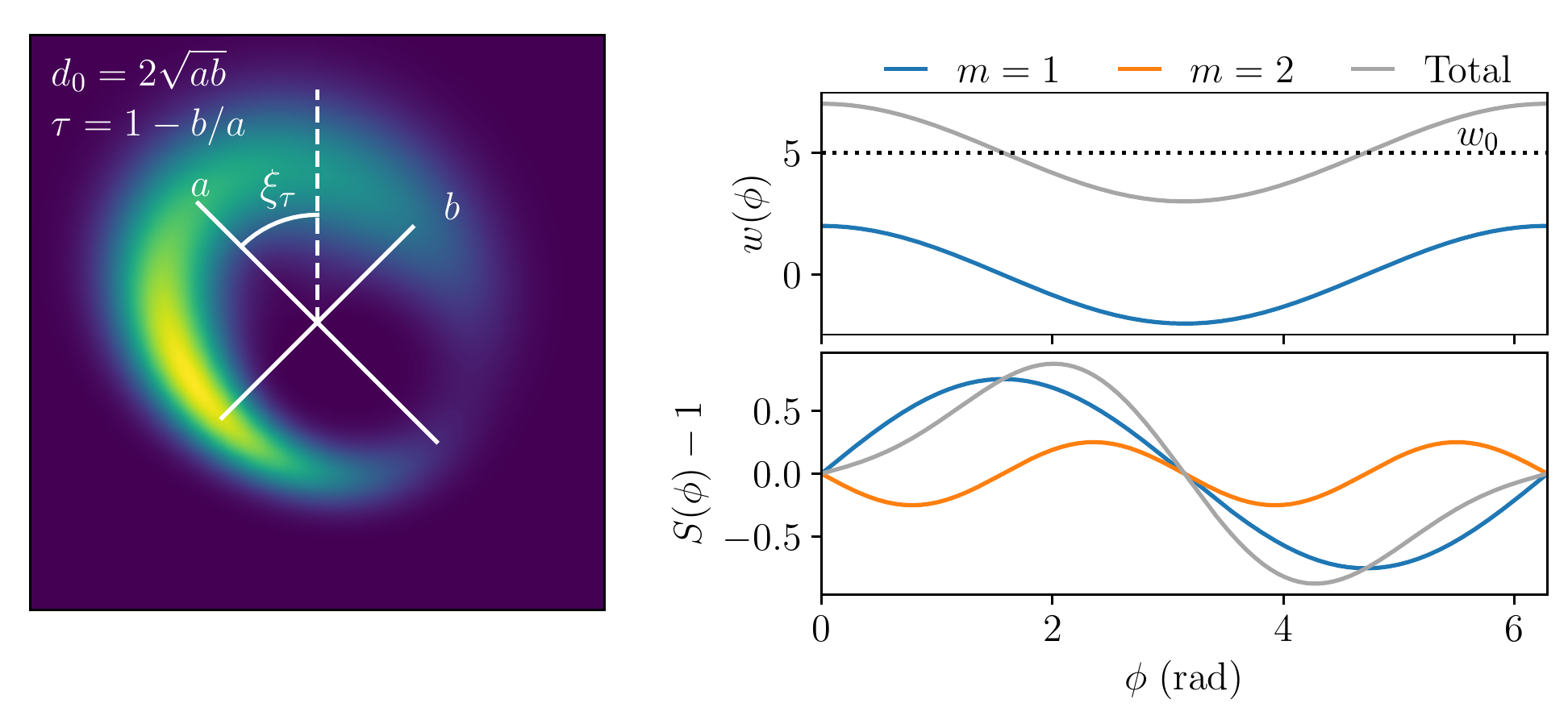}
    \caption{A visualization of the \VIDA \texttt{CosineRing}\{1, 2\} template with parameters $d0=42\muas$, $w_0 = 11.8\, \muas$, $w_1 = 4.7\muas$, $\xi_1^{(w)}=0$, $\tau = 0.2$, $\xi_\tau=\pi/4$,
    $s_1=0.75$, $s_2=-0.25$, $\xi_{1}^{(s)} = \pi/2$, $\xi_{2}^{(2)}=-\pi/4$. The left panel shows the visual appearance of the template normalized to have unit flux. The top right panel is the width profile and a function of azimuthal angle of the $N=1$ expansion in \autoref{eq:thickness_f}, where the black dotted line gives the ring width $w_0$. The bottom right is the brightness profile as a function of azimuthal angle of the $M=2$ expansion given in \autoref{eq:brightness}.}
    \label{fig:cosine_ring_cartoon}
\end{figure*}

The regularizers are important for the EHT, given its sparse coverage (see \autoref{fig:coverage}) and poor dynamic range \citepalias{EHTCII,EHTCIII}. Unfortunately, there is no canonical statistical framework\footnote{See \citet{akiyama_superresolution_2017} for an approach using cross validation that requires more data than currently available for the EHT} for how to choose the relative weights $\alpha_d$ and $\beta_r$. Instead, a series of heuristics and data quality metrics across a parameter survey of different regularizers are employed. In \citetalias{EHTCIV}, a survey of different hyperparameters was performed for each imaging method, DIFMAP, \texttt{SMILI}, and \texttt{eht-imaging} pipelines. This paper will focus on the \ehtim pipeline, but we found similar results for \texttt{SMILI}. The parameter surveys were run on a set of simple geometric models: a ring, a crescent, a disk, and a pair of small Gaussians to find an appropriate set of hyperparameters for the \m87 observations. These synthetic models were constructed with angular scales that approximately mimic the scale inferred from the EHT, thereby providing a test of whether pipelines can differentiate complex structure at equivalent angular scales.

Every combination of parameters in the survey was mapped to an effective resolution on each geometric model by interpolating normalized cross-correlations to the true image. The median performance across all \m87 observation days was used. This process enables comparisons of parameter performance across models; the effective resolution for each parameter combination was then averaged across models and used to rank all sets of parameters. The sets of parameters that produced an average effective resolution better than the EHT nominal resolution were used to form a ``top set.'' This top set is the effective set of parameters is that we will use to construct our ensembles of image reconstructions.

\subsection{Review of Feature Extraction Techniques}\label{sec:ftext}
\subsubsection{Variational Image Domain Analysis}\label{section:vida}

Variational image domain analysis (VIDA) \citep{vida} requires three ingredients: an image $I$, a template image $f_\theta$, and a probability divergence $\mathcal{D}$ that measures the difference between $f$ and $I$. \VIDA relies on the template function $f_\theta$ being a reasonable approximation to the true image. Given that we are interested in ring morphologies, we will use the \texttt{CosineRing}\{N, M\} template (see \autoref{fig:cosine_ring_cartoon} for a visualization)\footnote{For more information about the other templates present in \VIDA, please see \url{https://github.com/ptiede/VIDA.jl}}. The \texttt{CosineRing}\{N, M\} template is an elliptical Gaussian ring template, whose azimuthal brightness and thickness are described by a cosine expansion. More specifically, the model is described by:
\begin{itemize}
    \item diameter $d_0 = \sqrt{ab}$, where $a,\, b$ are the semi-minor and major axes respectively
    \item thickness function
    \begin{equation}\label{eq:thickness_f}
        w(\phi) = w_0 + \sum_{n=1}^{N}w_n \cos\left[n(\phi - \xi^{(w)}_m)\right]
    \end{equation}
    \item ellipticity $\tau = 1-b/a$, with orientation $\xi_\tau$ measured east of north
    \item slash function
    \begin{equation}\label{eq:brightness}
        S(\phi) = 1 - \sum_{m=1}^Ms_m\cos\left[m(\phi - \xi^{(s)}_m)\right]
    \end{equation}
    \item $x_0,\, y_0$, the center of the ring.
\end{itemize}
How to pick $N$ and $M$ is left to the user and the imaging problem at hand. \autoref{section:asym_cal_geom} is interested in recovering the profile for a simple elliptical ring with a slash. Therefore, we will take $N=0$ and $M=1$. We also tried higher-order mode expansions and found that they were typically much smaller than the first mode and did not change the results. For the GRMHD reconstructions in \autoref{section:asym_cal_grmhd} we take $N=1,\, M=4$, given their complicated azimuthal structure. This template has $16$ parameters in total. Typically, when we refer to the thickness or slash strength of the template, we will refer to the $w_0$ and $s_1$ parameters, respectively. In addition to the ring template, we add a constant intensity background where the background intensity is also a parameter. This intensity floor, models the diffuse intensity that is typically deposited across the image in reconstructions.

The optimal template is found by minimizing the divergence between the image reconstruction, normalized to unit flux, and the specified template. We will use the Bhattacharyya divergence, 
\begin{equation}
    \rm{Bh}(f_{\theta}||\mathcal{I}) = -\log\int \sqrt{f_\theta(x,y)\mathcal{I}(x,y)}\dd x \dd y,
\end{equation}
in all analyses below. To minimize the Bh divergence we use the Julia package BlackBoxOptim.jl\footnote{\url{https://github.com/robertfeldt/BlackBoxOptim.jl}} which uses an adaptive genetic algorithm for optimization. For more information on the validation of \VIDA and the optimization strategy see \citet{vida}.

\subsubsection{\ReX}
The other image feature extraction method used in this paper is the \textit{ring-extractor} or \ReX algorithm used in \citetalias{EHTCIV} and described in detail in \citet{ChaelThesis}. The first step in \ReX \citepalias[see][for details]{EHTCIV} is to identify the dominant ring in the image. First, for each pixel $x_i,y_j$ in the image, an intensity map $I(r, \theta |x_i, y_j)$ with radius and angle is defined relative to $x_i,y_j$. For each map, the ``radius'' of the image is given by:
\begin{equation}
    \begin{aligned}
        r_{\rm pk}(\theta|x,y) &= \mathrm{argmax}[I(r,\theta|x,y)]_r\\
        \bar{r}_{\rm pk} &= \left<r_{\rm pk}(\theta| x,y) \right>_{\theta \in [0,2\pi]}.
    \end{aligned}
\end{equation}
This provides a radius for each pixel center. The optimal center is then selected according to:
\begin{equation}
    (x_0,\; y_0) = \mathrm{argmin}\left[\frac{\sigma_{\bar{r}(x,y)}}{\bar{r}_{\rm pk}(x,y)}\right],
\end{equation}
where $\sigma_{\bar{r}}(x,y) = \left<(r_{\rm pk}(\theta|x,y) - \bar{r}_{\rm pk})^2\right>$, is the radial dispersion. This specifies the image center and all future quantities will be defined relative to this center.

\begin{figure*}[!t]
    \centering
    \includegraphics[width=0.8\textwidth]{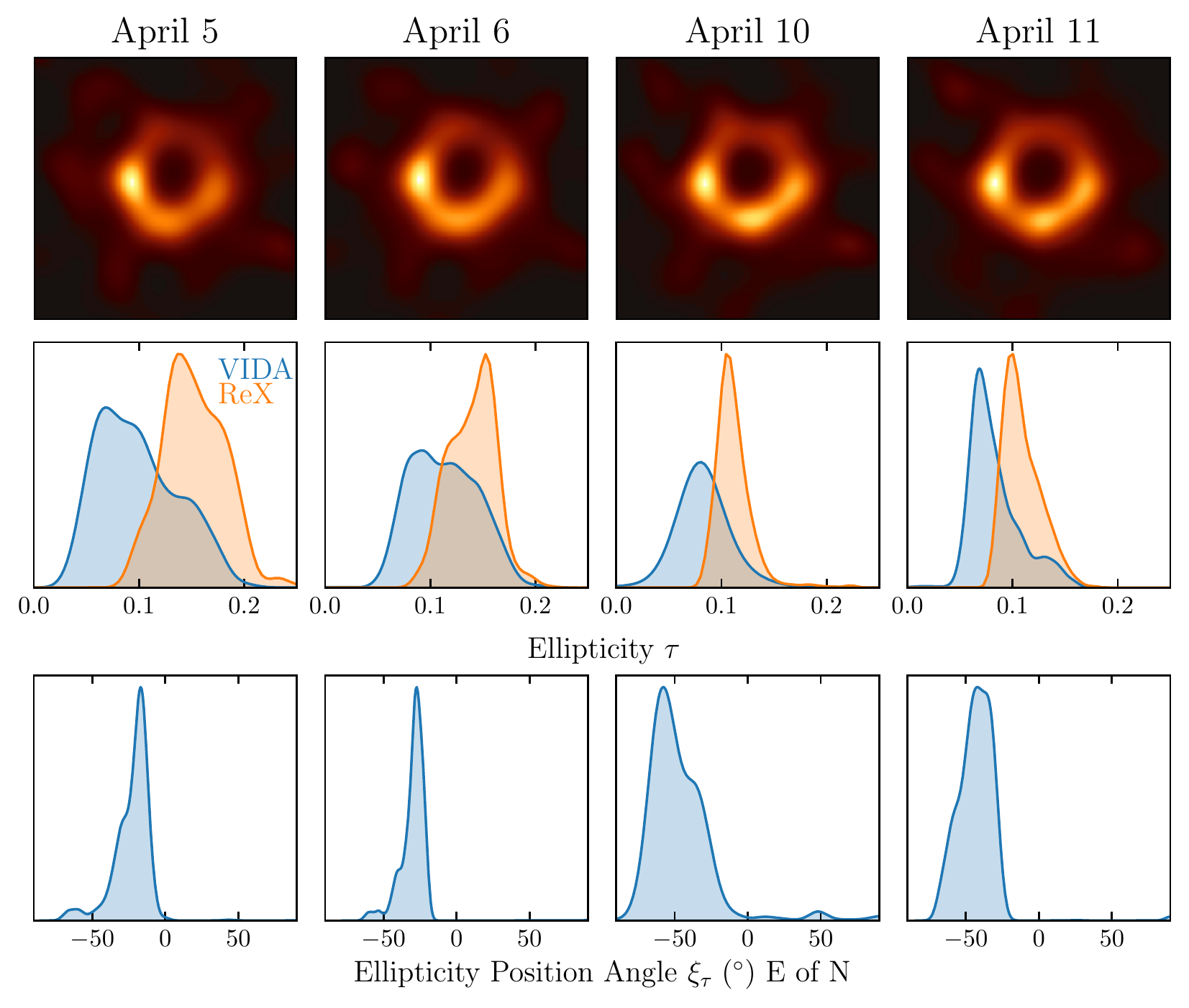}
    \caption{\ehtim reconstructions and ellipticity features of \m87 from the \citetalias{EHTCIV} top set across the different observations during the 2017 EHT campaign. The top row shows the fiducial \ehtim image from \citetalias{EHTCIV}. The middle row shows the measured ellipticity of \ehtim top set from \VIDA (blue) and \ReX (orange). The bottom row shows the orientation of the measured ellipticity orientation angle east of north from \VIDA. Note that \ReX is not currently able to measure the orientation angle. Overall we see that \m87 has consistent ellipticity around $\tau=0.05-0.2$, around $-50^\circ - 0^\circ$ east of north.}
    \label{fig:m87_asym_panel}
\end{figure*}

The diameter of the image is
\begin{equation}
    d = 2\bar{r}_{\rm pk}(x_0,y_0).
\end{equation}
Following \citet{EHTCVI}, \ReX characterizes the ellipticity of the ring structure by the radial fractional dispersion:
\begin{equation}\label{eq:rex_asym}
    f_d = \frac{\sigma_{\bar{r}}}{\bar{r}_{\rm pk}}.
\end{equation}
The width of the ring is defined by finding the full-width-half-max (FWHM) at a fixed $\theta$ ray, and then averaging over $\theta$,
\begin{equation}
    w = \left<\mathrm{FWHM}_r[I(r,\theta|x_0,y_0) - I_{\rm floor}]\right>_{\theta}.
\end{equation}
The intensity floor is given by $I_{\rm floor} = \left<I(r=50\muas, \theta)\right>_\theta$ and is included to avoid biasing the measurement due to the low level intensity present in the image. This is similar to including the constant intensity template during the \VIDA extraction.

To characterize the azimuthal profile of the ring ($\xi_s$ and $s$ for \VIDA) we consider the azimuthal moments of the ring. Similar to \VIDA we will only be interested in the first azimuthal moment. The orientation, $\xi_s$, of the first moment is given by:
\begin{equation}
    \xi_s = \left<\mathrm{Arg}\left[
            \int_0^{2\pi}I(r,\theta|x_0,y_0)e^{i\theta}\mathrm{d}\theta
            \right]       
            \right>_{r\in [r_{\rm in},r_{\rm out}]},
\end{equation}  
where $r_{\rm in} = (d-w)/2$ and $r_{\rm out} = (d+w)/2$. The coefficient of the first moment or slash is given by
\begin{equation}
    s = 2\left<\frac{\left|\int^{2\pi}_{0}I(r,\theta|x_0,y_0)e^{i\theta}\mathrm{d}\theta \right|}{\int^{2\pi}_{0}I(r,\theta|x_0,y_0)\mathrm{d}\theta}\right>.
\end{equation}

\begin{figure*}[!t]
    \centering
    \includegraphics[width=0.9\textwidth]{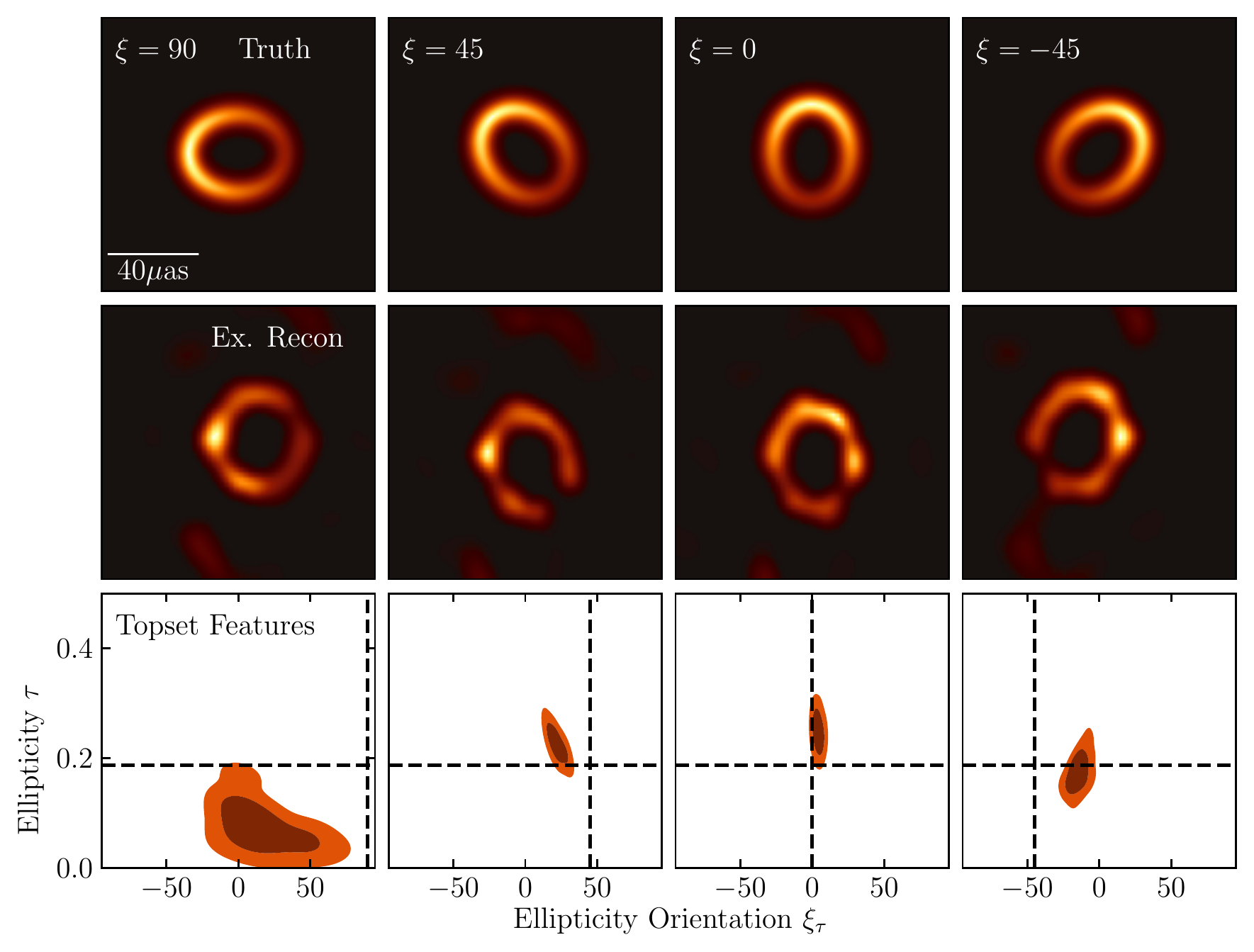}
    \caption{Examples of image reconstructions and \VIDA fits for the different elliptical rings. The top row shows the truth images at position angles $\xi_s = 0^\circ,\, 45^\circ,\, 90^\circ,\, 135^\circ$ north of east. The middle row shows an example reconstruction from the \m87 top set for each ring orientation. The bottom row shows the \VIDA results for the ellipticity and its positions angle from the top set. We found significant bias in $\tau$ and $\xi_\tau$ for rings whose semi-major axis was aligned in the east west direction.}
    \label{fig:ecc_examples}
\end{figure*}

\subsubsection{Relating \VIDA and \ReX Parameters}\label{sec:rexvida}
\VIDA and \ReX assume different parameterizations of the intrinsic structure, and use different optimization strategies. Therefore, we expect the resulting distributions to differ slightly. One prominent difference between \VIDA and \ReX is how they parameterize ellipticity. To compare both \VIDA and \ReX's ellipticity measurements, we need to relate \ReX's fractional dispersion $f_d$ to \VIDA's ellipticity $\tau$. To accomplish this we will convert \ReX's fractional diameter spread $f_d$ into \VIDA's $\tau$. First, we consider an ellipse with semi-major axis $a$ and semi-minor axis $b$. Then \VIDA parameterizes this ellipse with $d_0 = 2r_0 = 2\sqrt{ab}$ and $\tau = 1-b/a$. It can then be shown \citep[see ][Appendix A for a derivation]{vida} that the fractional dispersion is related to ellipticity $\tau$ through:
\begin{equation}\label{eq:rex_asym_conv}
    f_d(\tau) = \frac{\sqrt{1-\epsilon(\tau)^2 - 4/\pi^2 E^2\big(\epsilon(\tau)\big)}}{\sqrt{1-\tau}},
\end{equation}
where $E(x)$ is the complete Elliptic integral of the second kind and $\epsilon(\tau) = \sqrt{1-(1-\tau)^2}$ is the orbital eccentricity. Using linear interpolation we invert the function achieving a map from $f_d$ to $\tau$. 

This relation assumes that the image feature is a perfect ellipse. Given that \ReX defines the radius of the ring in terms of its peak, we expect that the shape will not be a perfect ellipse. This additional non-ellipticity will cause \ReX's ellipticity to be typically greater than \VIDA, which we empirically find below.
\begin{table*}[!ht]
\centering
\hspace{-25mm}
\begin{tabular}{lll| cccccc}
\hline\hline
 \textbf{Image} &&& $d_0$ & $w$ & $\tau$ & $\xi_\tau(^\circ)$ & $s$ & $\xi_s(^\circ)$ \\
\hline 
\multirow{2}{*}{$\xi_s=90^\circ$}&
\multirow{2}{*}{$\xi_\tau=90^\circ$}
        &\ReX 
        & $37.3^{+1.2}_{-1.0}$ 
        & $12.4^{+4.4}_{-4.4}$ 
        & $0.09^{+0.07}_{-0.04}$ 
        & \dots 
        & $0.54^{+0.07}_{-0.09}$ 
        & $90.68^{+3.9}_{-7.8}$ \\
        &&\VIDA 
        & $37.5^{+1.3}_{-1.2}$ 
        & $11.1^{+3.5}_{-4.0}$ 
        & $0.07^{+0.09}_{-0.05}$ 
        & $12.6_{-22.8}^{+48.8}$ 
        & $0.54^{+0.05}_{-0.07}$ 
        & $91.4^{+3.9}_{-10.4}$ \\
\cline{3-9}
\multirow{2}{*}{$\xi_s=45^\circ$}&
\multirow{2}{*}{$\xi_\tau=45^\circ$}
      &\ReX 
      & $37.2^{+0.8}_{-1.2}$ 
      & $11.2^{+4.7}_{-3.6}$ 
      & $0.20^{+0.04}_{-0.04}$ 
      & \dots 
      & $0.52^{+0.11}_{-0.12}$ 
      & $48.9^{+5.3}_{-7.8}$ \\
      &&\VIDA
      & $37.7^{+0.9}_{-1.4}$ 
      & $10.0^{+3.9}_{-3.4}$ 
      & $0.23^{+0.04}_{-0.03}$ 
      & $21.9_{-7.2}^{+8.1}$ 
      & $0.52^{+0.18}_{-0.12}$ 
      & $48.0^{+6.5}_{-9.8}$ \\
\cline{3-9}
\multirow{2}{*}{$\xi_s=0^\circ$}&
\multirow{2}{*}{$\xi_\tau=0^\circ$}
        &\ReX
        & $37.0^{+0.9}_{-1.3}$ 
        & $11.2^{+4.3}_{-3.7}$ 
        & $0.24^{+0.05}_{-0.03}$ 
        & \dots 
        & $0.45^{+0.15}_{-0.07}$ 
        & $-5.7^{+33.3}_{-20.6}$ \\
        &&\VIDA
        & $37.5^{+0.7}_{-1.3}$ 
        & $9.8^{+3.8}_{-3.3}$ 
        & $0.25^{+0.04}_{-0.04}$ 
        & $3.4_{-3.6}^{+5.3}$ 
        & $0.45^{+0.13}_{-0.07}$ 
        & $-3.4^{+27.1}_{-18.0}$ \\
\cline{3-9}
\multirow{2}{*}{$\xi_s=-45^\circ$}&
\multirow{2}{*}{$\xi_\tau=-45^\circ$} 
        &\ReX
        & $37.5^{+1.1}_{-1.1}$  
        & $11.7^{+4.3}_{-3.9}$ 
        & $0.18^{+0.04}_{-0.04}$  
        & \dots
        & $0.52^{+0.09}_{-0.09}$ 
        & $-49.4^{+22.1}_{-5.6}$  \\
        &&\VIDA
        & $38.3^{+0.9}_{-1.3}$  
        & $10.5^{+3.5}_{-3.5}$ 
        & $0.17^{+0.06}_{-0.05}$  
        & $-15.8_{-9.2}^{+8.8}$ 
        & $0.53^{+0.13}_{-0.04}$ 
        & $-50.7^{+19.8}_{-6.0}$  \\
\cline{3-9}
\multirow{2}{*}{$\xi_s=-90^\circ$}&
\multirow{2}{*}{$\xi_\tau=90^\circ$} 
        &\ReX 
        & $37.5^{+1.1}_{-1.2}$  
        & $12.4^{+4.4}_{-4.4}$ 
        & $0.09^{+0.07}_{-0.05}$  
        & \dots
        & $0.54^{+0.07}_{-0.12}$ 
        & $-90.3^{+6.4}_{-6.6}$  \\
        &&\VIDA 
        & $37.8^{+1.2}_{-1.2}$  
        & $11.1^{+3.5}_{-3.9}$ 
        & $0.06^{+0.09}_{-0.05}$  
        & $12.6_{-18.6}^{+42.4}$ 
        & $0.54^{+0.06}_{-0.08}$ 
        & $-90.0^{+6.2}_{-8.6}$  \\
\cline{3-9}
\multirow{2}{*}{$\xi=-135^\circ$}&
\multirow{2}{*}{$\xi_\tau=45^\circ$}
        &\ReX 
        & $37.1^{+0.7}_{-1.2}$  
        & $11.2^{+4.6}_{-3.6}$ 
        & $0.20^{+0.15}_{-0.03}$  
        & \dots
        & $0.47^{+0.12}_{-0.09}$ 
        & $-131.4^{+5.4}_{-6.0}$  \\
        &&\VIDA
        & $37.6^{+0.8}_{-1.2}$  
        & $9.9^{+3.8}_{-3.5}$ 
        & $0.22^{+0.04}_{-0.03}$  
        & $21.3_{-7.1}^{+8.8}$ 
        & $0.52^{+0.19}_{-0.09}$ 
        & $-134.4^{+4.4}_{-7.7}$  \\
\cline{3-9}
\multirow{2}{*}{$\xi=-180^\circ$}&
\multirow{2}{*}{$\xi_\tau=0^\circ$}
        &\ReX
        & $37.0^{+0.9}_{-1.3}$  
        & $11.1^{+4.2}_{-3.7}$ 
        & $0.24^{+0.04}_{-0.04}$  
        & \dots
        & $0.47^{+0.16}_{-0.08}$ 
        & $-184.1.2^{+20.4}_{-27.0}$  \\
        &&\VIDA
        & $37.6^{+0.8}_{-1.2}$  
        & $9.8^{+3.9}_{-3.3}$ 
        & $0.25^{+0.03}_{-0.03}$  
        & $4.3_{-3.5}^{+3.7}$ 
        & $0.49^{+0.16}_{-0.07}$ 
        & $-183.1^{+32.6}_{-12.9}$  \\
\cline{3-9}
\multirow{2}{*}{$\xi=135^\circ$}&
\multirow{2}{*}{$\xi_\tau=-45^\circ$}
        &\ReX
        & $37.6^{+1.2}_{-1.1}$  
        & $11.9^{+4.0}_{-4.3}$ 
        & $0.18^{+0.04}_{-0.03}$ 
        & \dots
        & $0.53^{+0.08}_{-0.05}$ 
        & $127.9^{+8.3}_{-5.3}$  \\
        &&\VIDA
        & $38.3^{+0.8}_{-1.2}$  
        & $10.5^{+3.9}_{-3.3}$ 
        & $0.17^{+0.03}_{-0.03}$  
        & $-13.2_{-3.5}^{+3.7}$ 
        & $0.56^{+0.16}_{-0.07}$ 
        & $127.9^{+7.5}_{-6.1}$  \\
\hline
\multicolumn{1}{l}{\textbf{Truth}} &&& 37.56 & 7.9 & 0.187 & \dots & 0.5 & \dots \\
\end{tabular}
\caption{Recovered parameters for the slashed elliptical rings test set. The parameters are the median values and the 95\% interval around the median. \ReX and \VIDA give very similar results for all parameters, although no results for $\xi_\tau$ are given for \ReX since it cannot recover it. All the parameters except the ellipticity $\tau$ and its orientation $\xi_\tau$ contain the true values.}\label{table:asym_ex}
\end{table*}

\subsection{Review of \m87 Ellipticity Measurement}
Figure 18 of \citetalias{EHTCVI} showed the measured \m87 fractional deviation from \ReX applied to the \ehtim top set parameters. To reproduce these results, we applied \VIDA and \ReX to the \ehtim top set. The results are shown in \autoref{fig:m87_asym_panel}. The top row shows the fiducial image reconstruction from the \ehtim top set across each observation day. The ellipticity of the top set images is shown in the middle row. We find identical results to those in \citetalias{EHTCVI} for \ReX. The \VIDA results are systematically lower than \ReX, as expected from the discussion in the previous section. The bottom row presents, for the first time, the orientation of this ellipticity from \VIDA. Note that \ReX cannot currently measure this orientation. Overall, we find that the ellipticity measurement is stable across all four days, giving $\tau=0.05-0.2$, and orientation $\xi_\tau=-75^\circ-0^\circ$ east of north.

While it is interesting that the ellipticity measurements are consistent across days, it is not clear whether this result is intrinsic to the source. In \citet{vida}, we found statistically identical ellipticity and orientation for the symmetric crescent models and GRMHD models. Given that the crescent models are symmetric, it suggests that the measured ellipticity may be an imaging artifact. Furthermore, the ellipticity orientation does align with a coverage gap (see the red circles in \autoref{fig:coverage}). Finally, since the top set used for \m87 did not include an elliptical ring, it is not clear whether ellipticity can reliably be recovered. To investigate the fidelity of ellipticity reconstruction using the reported top set, we will apply it to simple elliptical ring images in the next section.

\section{Geometric Test}\label{section:asym_cal_geom}
One of the potential issues with the \m87 top set is that no elliptical rings were included in the simulated data tests. Given that the identification of the top set was defined by its performance on simulated data tests, the top set may not accurately recover ring ellipticity even in simple cases. This section will analyze simulated data from an elliptical ring model using the \m87 top set.

\subsection{Elliptical Image Test}
For the elliptical image, we used the \texttt{CosineRing}\{0,1\} template described in \autoref{section:vida}, with parameters $d_0=37.56\, \muas$, $w = 7.9\, \muas$, $\tau = 0.187$, $s = 0.5$. The ring flux was set $0.6\, \rm Jy$, matching the measured compact flux of \m87. We also aligned the orientation of the slash and ellipse, i.e., we set $\xi_s=\xi_\tau = \xi$. To test the impact of different orientations of the ellipticity we considered $\xi = 0^\circ$ to $360^{\circ}$ in steps of $45^\circ$. A subset of the ground truth images are shown in the top row of \autoref{fig:ecc_examples}. For each rotated ring we used the \ehtim top set pipeline from \citetalias{EHTCIV} to create 1572 reconstructions.

\subsection{Geometric Results}
Given the elliptical ring reconstructions, we used \VIDA and \ReX to extract the relevant image features. Since the \VIDA template is identical to the on-sky image, we expect that the true parameter value will lie within the distribution of recovered image features. The results for each orientation are summarized in \autoref{table:asym_ex}. The ring diameter and width are consistent across the rotation angles, and are consistent with the truth. The slash and its orientation are similarly recovered. 

However, the ellipticity, $\tau$, is significantly biased when $\xi_\tau = 90^\circ$, i.e., when the semi-major axis of the ellipse is aligned in the east-west direction. Furthermore, looking at the bottom row of \autoref{fig:ecc_examples}, we see that the orientation of the ellipticity is consistently biased towards $\xi_\tau = 0$, i.e., the north-south direction. This bias can be visually confirmed by looking at the reconstructions, e.g., the middle row of \autoref{fig:ecc_examples}. Furthermore, we see that the true ellipticity and orientation is only recovered in the $\xi=0$ case.

Therefore, it appears that imaging creates a preferred ellipticity direction  $\xi_\tau\approx 0$, $\tau$. Namely, as $\tau$ increases, the ellipticity orientation tends to point in the north-south direction. This orientation does approximately align with a large gap in the EHT coverage for \m87 (see the red circles in  \autoref{fig:coverage}).

Given the inability of the top set images to faithfully recover the ellipticity in our geometric tests, it is not clear that the measurement found in \autoref{fig:m87_asym_panel} represents a constraint on the intrinsic ellipticity and not simply an artifact of the top set itself.  To address this possibility, we will calibrate the procedure applied above to a large number of GRMHD simulations.

\section{Calibrating the \m87 Ellipticity Measurement}\label{section:asym_cal_grmhd}
To calibrate for the uncertainty in the \m87 top set ellipticity, we will use a similar procedure to the mass calibration done in \citetalias{EHTCVI}. First, we selected a number of GRMHD simulations from the \citetalias{EHTCV} library to provide physically motivated images. Second, we constructed simulated data matching the 2017 \m87 observations and reconstruct the images using the \m87 \ehtim top set. Using this set of reconstructions, we measured the ellipticity of each reconstruction and compared it to the ground truth constructing a ``theoretical'' uncertainty budget for the ellipticity. This uncertainty is then included in the uncertainty of \m87's ellipticity.

\subsection{Scaled set}\label{section:grmhd_scaled}
To construct the GRMHD images used in this paper, we first cut the simulations from \citetalias{EHTCV} based on whether its total jet power was consistent with the observed jet power of M87 \citep[Table 2][]{EHTCV}. From the remaining set of simulations, we randomly select 100 snapshots and randomly assign them to a 2107 \m87 observation day. Each selected snapshot is re-scaled to its best fit value (according to the average image scoring results of \citet[][]{EHTCVI}), and randomly rotated. To include the effects of the mass uncertainty of \m87, we then further scaled the intrinsic image by a factor of 0.8, 0.9, 1.1, 1.2 in both the x and y directions. The net result is 500 images uniformly sampled over days, orientations and grided in mass relative to the \m87 best fit value. We will refer to this list as the \textit{scaled set}.

\subsection{Stretched set}\label{section:grmhd_stretched}
While the scaled set measures the expected ellipticity due to imaging and accretion turbulence, it does not measure how sensitive imaging is to additional intrinsic ellipticity that may occur from, e.g., non-GR spacetimes. To assess the ability to measure an elliptical shadow, we randomly selected 10 additional GRMHD snapshots that fit M87's jet power. For each image, we scaled them to their best fit mass and randomly rotated them. Ellipticity was added by picking two random orthogonal directions in the image $r_x,\, r_y$ and applying the transformation $r_x\to\alpha r_x$ and $r_y \to r_y/\alpha$. To create different amounts of ellipticity we let $\alpha=0.8,0.9,1.0,1.1,1.2$, giving a $\tau = 0.36, 0.19, 0.0, 0.17, 0.31$ respectively. We will refer to this as the \textit{stretched set} in the remainder of the paper.

\begin{figure*}[!ht]
\centering
\includegraphics[width=0.42\linewidth]{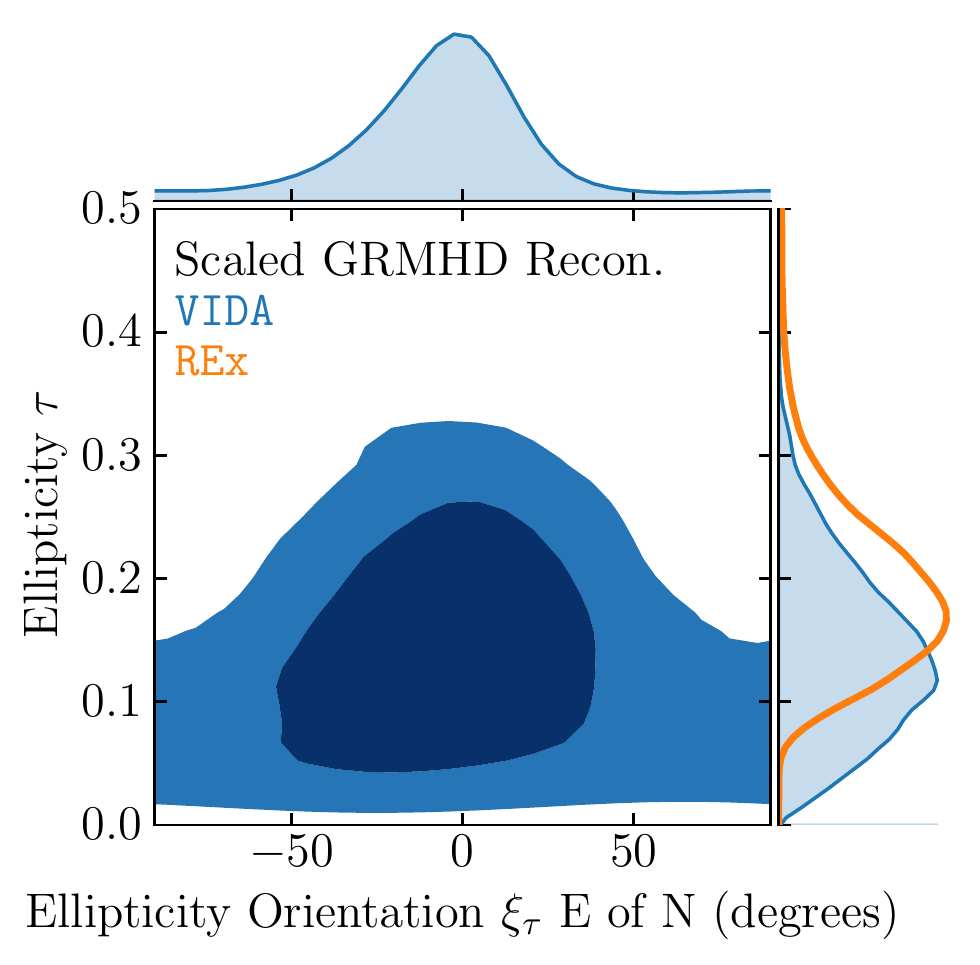}
\includegraphics[width=0.42\linewidth]{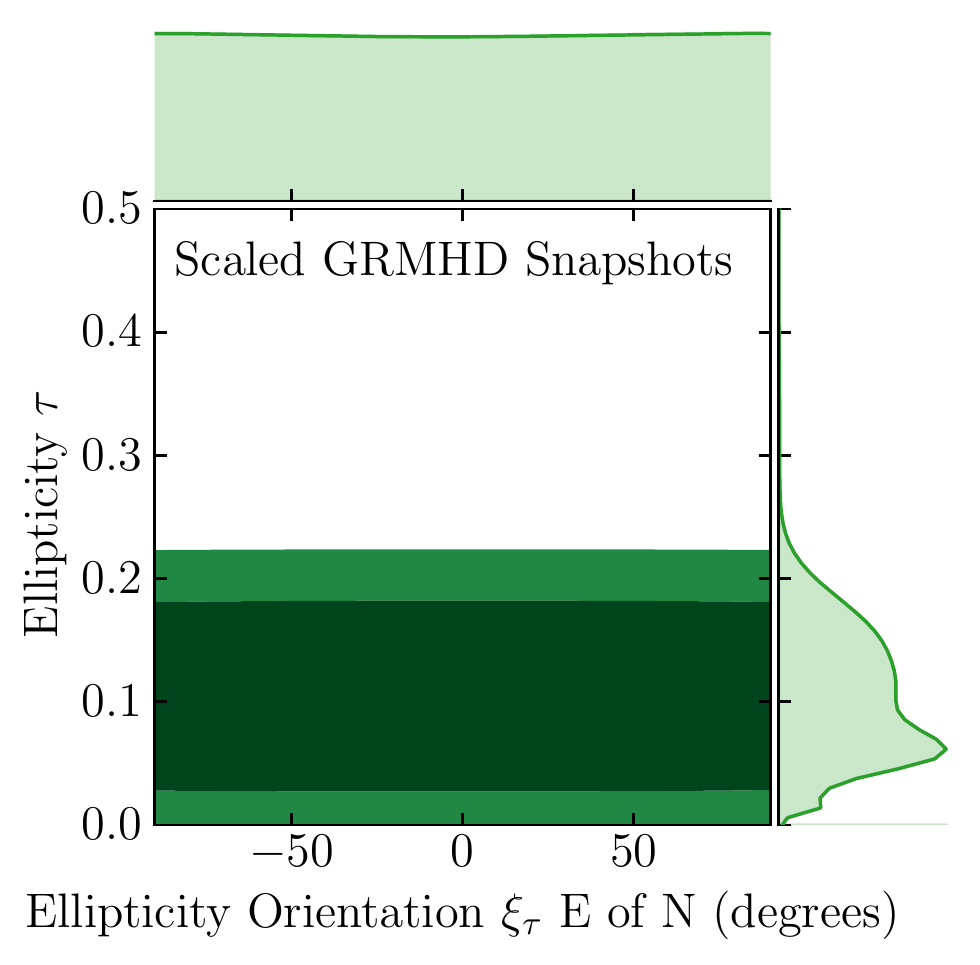}\\
\includegraphics[width=0.42\linewidth]{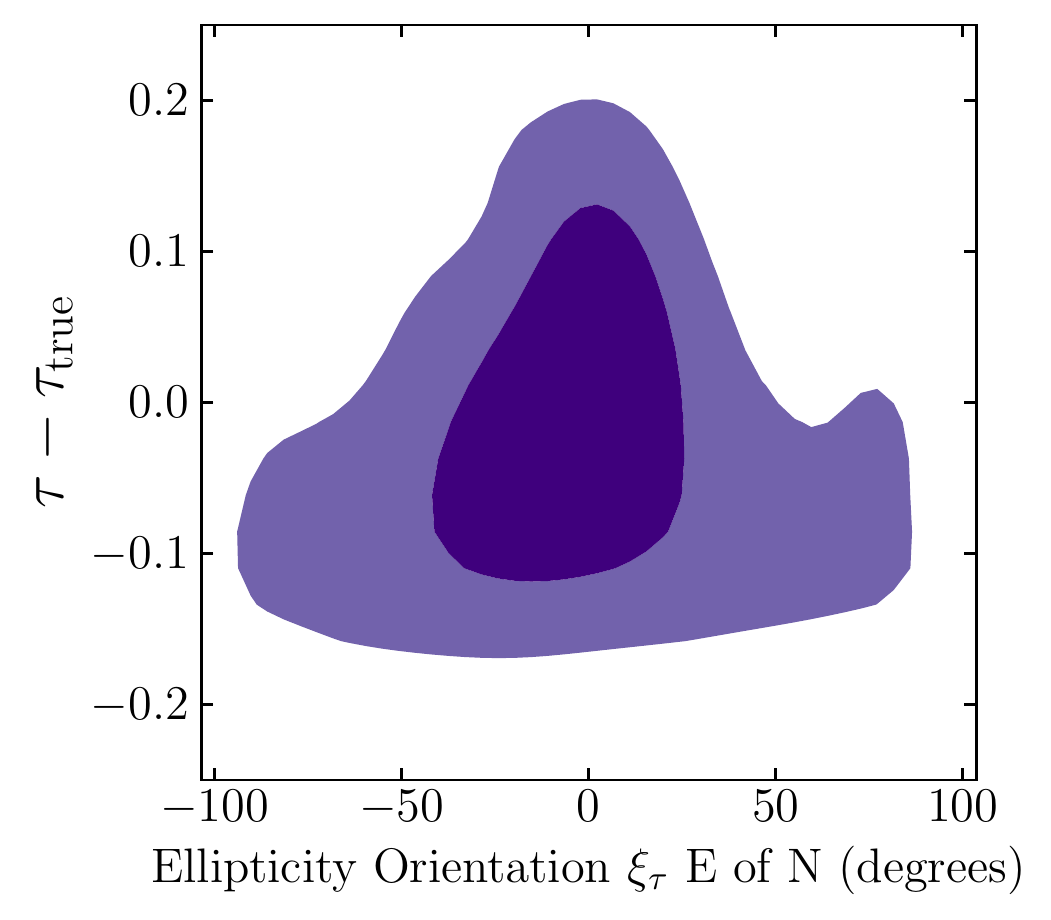}
\caption{Joint marginal distribution between $\tau$ and $\xi_\tau$, where the contours are the $68\%$ and $95\%$ regions. The upper left shows the results for the entire reconstructed scaled GRMHD set that satisfies $a_{\rm waffle} < a_{\rm thresh}=0.1$ threshold. The upper right is when \VIDA is applied directly to the GRMHD snapshot blurred with a $15\,\muas$ FWHM Gaussian kernel. Like the geometric results, we see a preference for $\xi_\tau\approx 0^\circ$, i.e., the north-south direction regardless of the intrinsic image distribution. The bottom figure shows the measured ellipticity orientation on the x-axis with the measured ellipticity residual distribution.}\label{fig:joint_marg_scaled}
\end{figure*}

\subsection{Removing failed top set reconstructions}\label{section:classify}
While inspecting the top set reconstructions of the scaled and stretched GRMHD simulations, we noticed that a large number of images failed to show a ring-like feature. Instead, the image reconstructions had intensity deposited across the image in a pattern similar to the EHT dirty beam. This is commonly known as ``waffling'' and is symptomatic of a poorly chosen set of hyperparameters. For these reconstructions, \VIDA and \ReX would give nonsensical results since no dominant ring feature exists. To remove this bias, these reconstructions need to be removed. Unfortunately, no single set of hyperparameters was identified as having caused the waffling. Therefore, we turned to machine learning techniques to remove any waffled images.

We created a deep convolutional neural network (CNN) implemented in \texttt{Flux.jl} \citep{Flux.jl-2018} to classify and remove the waffled images. More details about the network and image classification are given in \autoref{appendix:cnn}. The trained neural network outputs a number $\gamma$ between 0 and 1 that measures the confidence that the image has waffled. We then cut any image reconstructions where $\gamma > \gamma_{\rm thresh} = 0.1$. This threshold cut $42\%$ of images in the scaled set and 21\% of the images in the stretched set. The impact of the value of $\gamma_{\rm thresh}$ is shown in \autoref{fig:asym_threshold}. As an additional test of the network, we ran the classifier on the elliptical Gaussian reconstructions. We found that only $\sim$5\% of the elliptical images were cut, which is consistent with our visual inspection.

\begin{figure}[!h]
    \centering
    \includegraphics[width=1.0\linewidth]{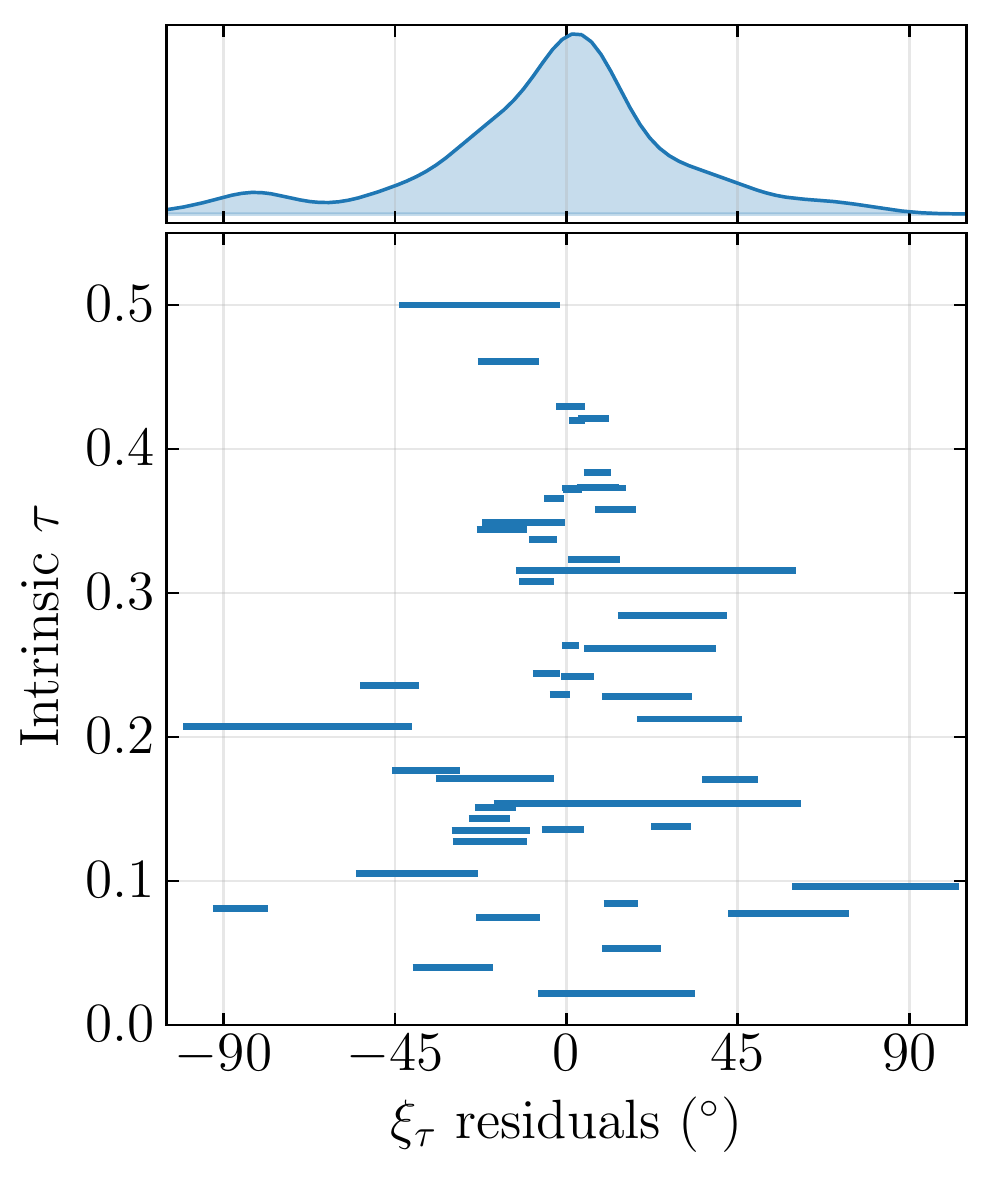}
    \caption{Residual distributions of the stretched GRMHD data sets for the ellipticity positions angle $\xi_\tau$. The top panel shows the marginal distribution of the ellipticity residual across the entire stretched GRMHD set. The bottom shows the $95\%$ probability interval of the position angle residuals for each simulation, separated by the intrinsic $\tau$ on the y-axis. The intrinsic $\tau$ and $\xi_\tau$ were found by applying \VIDA to the GRMHD snapshots blurred by a Gaussian with a FWHM of $10\muas$. The large residuals for small $\tau$ are a result of the $\xi_\tau$ being heavily biased north-south, similar to the elliptical ring, and scaled GRMHD results.}
    \label{fig:stretched_res}
\end{figure}

\subsection{Scaled set results}
We used \VIDA's \texttt{CosineRing\{1,4\}} template and Bh divergence to analyze the scaled set of images. Using the CNN image classifier (\autoref{section:classify}) we removed the ``bad'' reconstructions, leaving 454,888 images. To compare the results to the GRMHD simulation's ellipticity, we first blurred the ground truth snapshots with a Gaussian kernel with FWHM $15\muas$ to model the finite resolution of the EHT array. Then we fit the blurred images with the \VIDA and the \texttt{CosineRing\{1,4\}} template. This formed our ground-truth ellipticity, for which we will compare all results below.

The results for $\xi_\tau$ and ellipticity $\tau$ for the entire scaled set is shown in the upper left panel of \autoref{fig:joint_marg_scaled}. We found that the ellipticity is quite uncertain in both \VIDA and \ReX extending to $\tau=0.3$, which is approximately a $5:3$ axis ratio. This ellipticity is larger than ground truth ellipticity measured from \VIDA (right panel of \autoref{fig:joint_marg_scaled}), which is quite concentrated at $\tau\approx 0.1$ and extends up to $\tau=0.25$. As mentioned in \autoref{sec:intro}, the spin of the black hole is expected to add a small amount of ellipticity to the on-sky image. We found no evidence for any ellipticity-spin correlation. The lack of correlation provides evidence that the dominant source of ellipticity in the reconstructions comes from the accretion flow itself.

Focusing on the ellipticity orientation $\xi_\tau$, we find the ellipticity is strongly biased in the north-south direction. Furthermore, when $\xi_\tau\approx 90^\circ$, $\tau$ extends to highest values of ellipticity. This orientation distribution is inconsistent with the GRMHD distribution (right \autoref{fig:joint_marg_scaled}), which is uniform in $\xi_\tau$. The uniform distribution was the expected result since each simulation is randomly rotated before imaging. The north-south bias is similar to the results found for the elliptical ring in \autoref{section:asym_cal_geom}, and the circular crescents in \autoref{fig:m87_asym_panel}.

Taking the measured ellipticity and orientation bias together suggests that the imaging algorithms create a preferred ellipticity orientation, and along this direction, the ellipticity uncertainty is maximized. This is shown in the bottom panel of \autoref{fig:joint_marg_scaled}. Here we see that when the ellipticity orientation is aligned in the N-S direction, the recovered ellipticity is very uncertain and can be quite different from the truth.  Additionally, for $\xi_\tau$ approaching $\pm 90^\circ$, the top set tends to prefer overly circular images. 

To add this ellipticity uncertainty to the results for \m87 we take each recovered ellipticity and orientation from the \m87 top set and add the theoretical uncertainty. This gives that the ellipticity of \m87 is $\tau \in [0.0, 0.3]$. Note this result assumes the accretion flow around \m87 is well described by a GRMHD simulation, and there are no non-Kerr effects that add ellipticity to the image. In the next section, we will analyze what happens for the stretched set of GRMHD simulations that include additional ellipticity.

\begin{figure*}[!ht]
    \centering
    \includegraphics[width=\linewidth]{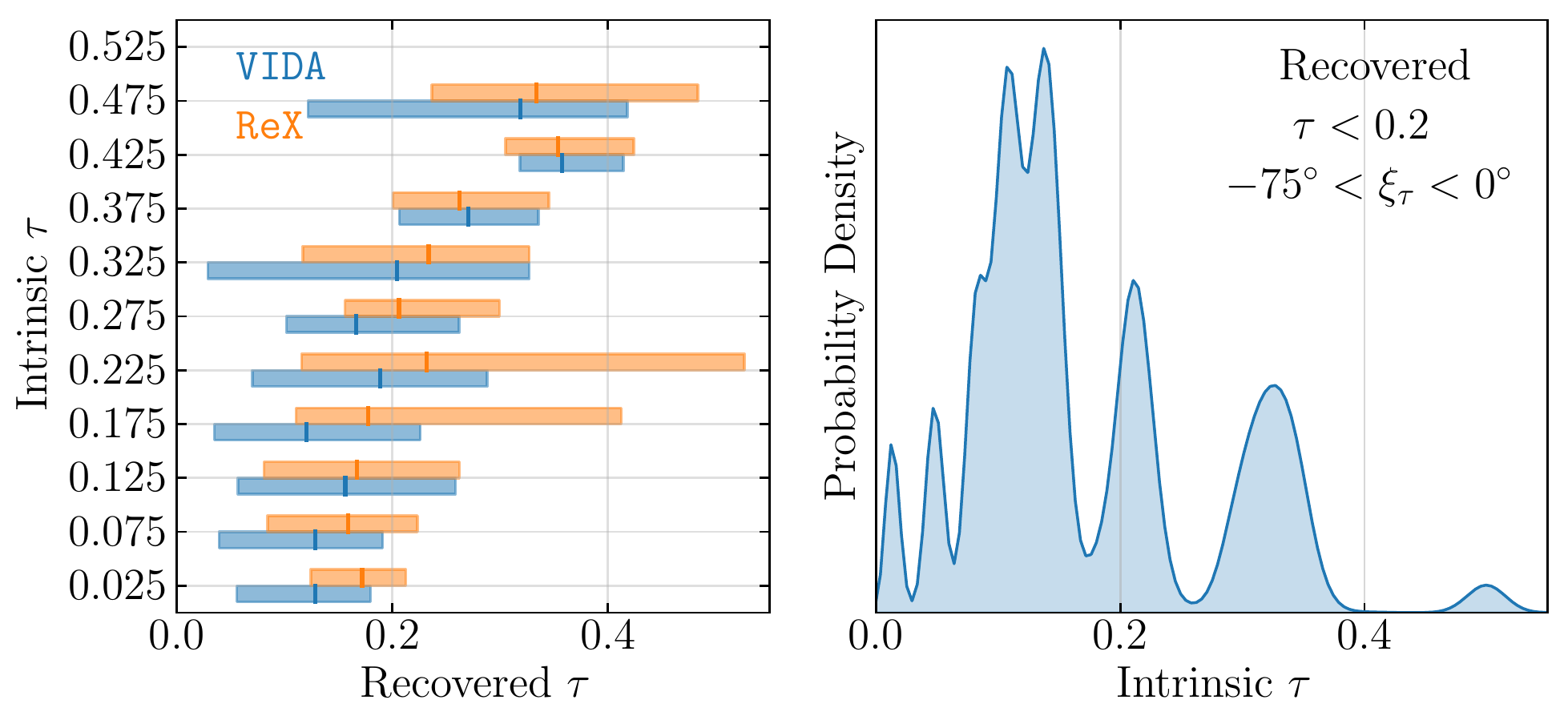}
    \caption{Left: Intrinsic ellipticity $\tau$ vs recovered ellipticity of the stretched GRMHD images. The intrinsic ellipticity was found by applying \VIDA to the GRMHD snapshot images blurred with a $15\muas$ Gaussian kernel. The solid line shows the median recovered ellipticity from \VIDA (blue) and \ReX (orange). The filled-in bars are the $95\%$ confidence intervals around the median. Right: The resulting intrinsic ellipticity distribution (measured by \VIDA), for reconstructions with reconstructed ellipticity and orientation similar to the observed \m87 top set values.}\label{fig:stretch_corr}
\end{figure*}

\subsection{Stretched set results}
While the results from the scaled set of GRMHD simulation suggest an upper bound of $\tau\lesssim 0.35$, it does not answer what happens when there is additional ellipticity, not due to the accretion disk. For instance, non-GR metrics \citep[e.g.]{Johannsen2010} can cause the black hole photon ring and/or event horizon to appear more elliptical than what the Kerr metric would predict. To test this, we will use the GRMHD stretched data set described above. To extract ellipticity and orientation from the stretched set, we again used \VIDA's \texttt{CosineRing\{1,4\}} template with a constant background and the Bh divergence. However, we found that a small subset of the image reconstructions had an additional circular blob present. Due to the second component, \VIDA would sometimes report an artificially high ellipticity since the ring template would try to cover both the central ring and Gaussian blob. To remedy this issue, we added a Gaussian component to the template to model the non-ring intensity. We found that this worked as expected and the ring template gave a more reliable ellipticity measurement.

The residuals for the ellipticity position angle are shown in \autoref{fig:stretched_res}. Note that we only show the results for \VIDA since \ReX cannot measure the ellipticity's position angle. Looking at the bottom panel, we see that for intrinsic $\tau \lesssim 0.2$, the position angle residuals are very broad and can be significantly biased from zero. However, as the intrinsic $\tau$ increases we find that the residuals improve. This bias for smaller ellipticity is due to $\xi_\tau$ being strongly biased towards $0^\circ$, and therefore away from a small residual. This bias continues as $\tau$ increases, but its impact is lessened.

The left panel of \autoref{fig:stretch_corr} shows a map from the recovered $\tau$ on the x-axis and the intrinsic $\tau$. The horizontal bars are the $95\%$ confidence regions about the median of the top set images. From this, we see that the recovered ellipticity is largely independent of the intrinsic ellipticity when the intrinsic $\tau \lesssim 0.325$. Furthermore, even a GRMHD simulation with intrinsic $\tau=0.475$ can have a recovered $\tau=0.1$, which is similar to the value obtained for the \m87 top set. 

To quantify the maximum allowed ellipticity that is consistent with \m87, we first made two cuts on the stretched GRMHD reconstructions. Namely, we remove any reconstructions where $\tau > 0.2$ and $\xi_\tau \notin \left[-75^\circ, 0^\circ\right]$. The remaining simulations, therefore, match the \m87 top set measurements. The intrinsic ellipticity distribution from the remaining simulations is shown in the right panel of \autoref{fig:stretch_corr}. This distribution shows that stretched simulations with ellipticity as high as $0.5$ can have ellipticity similar to the observed \m87 results. Note that this is the highest intrinsic ellipticity considered in the stretched set.

\section{Summary and Conclusions}\label{sec:conclusions}
The ellipticity of the accretion flow around \m87 is a theoretically interesting property related to the nature of the accretion flow and structure of spacetime. While the results in \citetalias{EHTCVI} measured an ellipticity in the image reconstruction of \m87, no attempt was made to interpret or calibrate this result. However, we have shown that the top set used for the \m87 images cannot directly measure the on-sky image ellipticity. We demonstrated that the \ehtim top set failed to recover the correct ellipticity in 8/10 test cases, even for simple geometric models.

To account for ellipticity bias, we calibrated the \m87 ellipticity using a set of 550 GRMHD images. Assuming that the ellipticity in the reconstructions is due to accretion turbulence, we found that accounting for the imaging bias, the ellipticity of \m87 could be anywhere from $\tau=0$ to $\tau=0.3$. However, if there is additional non-accretion induced ellipticity from, for example, some non-GR effect, we found that ellipticity as high as $\tau \approx 0.5$ could have a recovered ellipticity similar to the \m87 results.

The reason for the ellipticity uncertainty is twofold. First, the uv coverage for the EHT 2017 data is very sparse, having significant gaps in the north-south direction. Secondly, the top set used for \m87 is inadequate for both the geometric and GRMHD simulations analyzed in this paper. This can be seen from the strong north-south ellipticity bias. Further evidence of the top set inadequacy comes from the roughly $20-40\%$ of the reconstructed images that needed to be removed from the results as detailed in \autoref{section:classify}.

The results of this paper also stress the importance of defining parameter surveys that include image features that are of interest. If the survey does not include the impact of these image features, it is not clear whether the resulting reconstruction measurements of those features will be reliable predictors of the true on-sky image. This partially requires some preliminary understanding of the true on-sky image features. Unfortunately, this is often unclear for the EHT, making designing parameter surveys difficult. Fundamentally, the reason for this difficulty is that while the top set provides an ensemble of image reconstructions, they are decidedly not a posterior over image structures. Instead, they rely on training sets to decide which set of images meets some heuristic threshold. The top set is, therefore, unable to measure image feature uncertainty. Bayesian imaging techniques are needed to measure the image ellipticity posterior. In future work, we will apply the Bayesian imaging methods in \citet{Broderick2020} to measure the image ellipticity statistically.

Future observations will significantly reduce the ellipticity uncertainty and disentangle the different potential causes of any measured ellipticity and increase our knowledge of black hole parameters and accretion \citep{Roelofs2021}. The next-generation EHT will potentially add ten additional sites across the globe, significantly increasing coverage and dynamic range. By improving the dynamic range, the detection of the inner shadow with the ngEHT \citep{Doeleman_ngEHT, Raymond} may be possible \citep{ChaelInnerShadow}. The improved coverage will also reduce the need for ellipticity calibration that was required due to image artifacts. Proposed space-based arrays like the \emph{Event Horizon Explorer} \citep{EHE} will also have the potential to significantly deepen our understanding of accretion \citep{Leonid} directly model or image the $n=1$ photon ring \citep{Gralla2020, Paugnat2022} and constrain its shape.

\begin{acknowledgments}

P.T. receives support from the Natural Science and Engineering Research Council through the Alexander Graham Bell CGS-D scholarship. 
This work was made possible by the facilities of the Shared Hierarchical Academic Research Computing Network (SHARCNET:www.sharcnet.ca) and Compute/Calcul Canada (www.computecanada.ca).
Computations were made on the supercomputer Mammouth Parall\`ele 2 from University of Sherbrooke, managed by Calcul Qu\'ebec and Compute Canada. The operation of this supercomputer is funded by the Canada Foundation for Innovation (CFI), the minist\`ere de l'\'Economie, de la science et de l'innovation du Qu\'ebec (MESI), and the Fonds de recherche du Qu\'ebec - Nature et technologies (FRQ-NT).
This work was supported in part by Perimeter Institute for Theoretical Physics.  Research at Perimeter Institute is supported by the Government of Canada through the Department of Innovation, Science and Economic Development Canada and by the Province of Ontario through the Ministry of Economic Development, Job Creation and Trade.
A.E.B. thanks the Delaney Family for their generous financial support via the Delaney Family John A. Wheeler Chair at Perimeter Institute.
A.E.B. and P.T. receive additional financial support from the Natural Sciences and Engineering Research Council of Canada through a Discovery Grant. D.C.M.P. and P.T. was supported by the Black Hole Initiative at Harvard University, which is funded by grants from the John Templeton Foundation and the Gordon and Betty
Moore Foundation to Harvard University. D.C.M.P. and P.T. were also supported by National Science Foundation grants AST 19-35980 and AST 20-34306. A.C. is supported by Hubble Fellowship grant HST-HF2-51431.001-A awarded by the Space Telescope Science Institute, which is operated by the Association of Universities for Research in Astronomy, Inc., for NASA, under contract NAS5-26555.

\end{acknowledgments}

\software{BlackBoxOptim.jl \citep{BlackBoxOptim}, \texttt{eht-imaging} \citep{Chael_2018}, Flux.jl \citep{Flux}, GR \citep{gr}, Julia \citep{julia}, matplotlib 3.3 \citep{matplotlib}, Pandas \citep{pandas}, Python 3.8.3 \citep{python3}, Scipy \citep{SciPy}, ThemisPy \citep{ThemisPy}, VIDA.jl \citep{vida}}

\clearpage

\bibliographystyle{aasjournal}
\bibliography{references}

\appendix
\section{Image Reconstruction classification using CNNs}\label{appendix:cnn}
\begin{figure}[!hb]
    \centering
    \includegraphics[width=\linewidth]{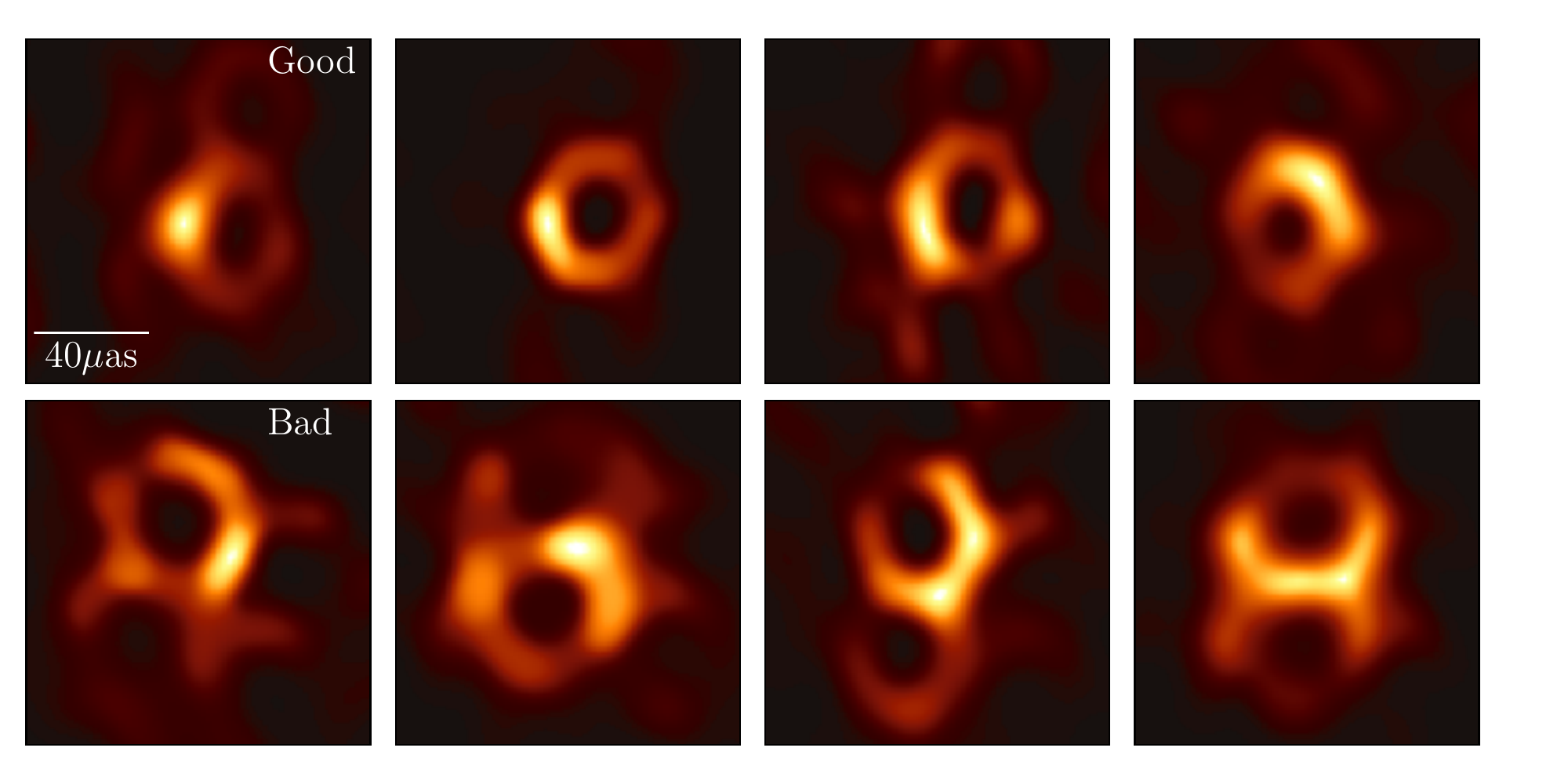}
    \caption{Examples of image classification from the trained CNN. The top row shows image reconstructions that passed the threshold, and the bottom are ones that failed. Most of the failed reconstructions fail to demonstrate a single dominant ring-like feature in the image.}
    \label{fig:waffler}
\end{figure}
While analyzing the image reconstructions of the GRMHD simulations, we found that a significant portion of the images failed to show a dominant ring-like feature. This failure can occur when the training set of images used in the top set's construction is different from the true image (e.g., the scaled GRMHD images). When this occurs, the reconstruction will appear to ``waffle.'' The waffling results in flux being spread throughout the imaging plane, in a pattern similar to the EHT's dirty beam. Unfortunately, no global section of the top set was able to remove the waffled images. Therefore, we turned to machine learning techniques to classify the reconstructions as having succeeded or failed.

Machine learning and neural networks have been used numerous times in astrophysical settings to classify images \citep[e.g.]{Davelaar2017,dieleman_rotation-invariant_2015}. To classify our images we decided to use convolutional neural networks (CNNs) \citep[e.g.][]{Goodfellow2016}. CNNs break the images into features of different scales and then group these features to classify the image. For our neural network, we used the Julia package Flux \citep{Flux.jl-2018,innes:2018}.

\begin{figure}[!t]
    \centering
    \includegraphics[width=0.5\linewidth]{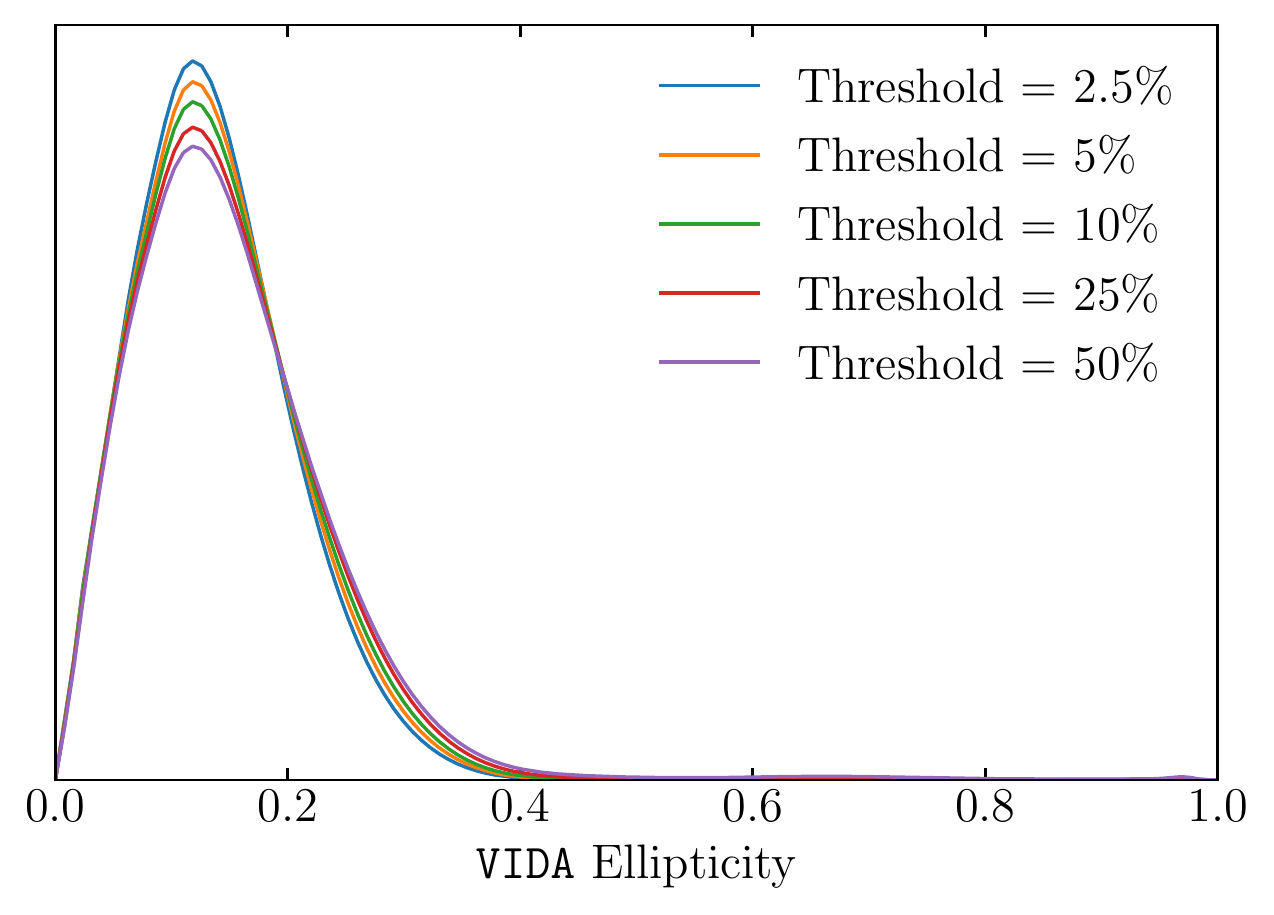}
    \caption{Impact of the threshold on the inferred ellipticity of the scaled reconstructions. Overall there is very little change in the distribution if the threshold is $\leq 0.5$. In this paper we use the threshold of $0.10$ which removes $42\%$ of the scaled reconstructions.}
    \label{fig:asym_threshold}
\end{figure}

For our image classifier, we used a relatively shallow network using 3 convolutional layers with a $2\times 2$ max pooling. For the final layer, we used a fully connected network to the two-dimensional classification space. Between each convolutional layer we used the ReLu activation function: $f(x) = \mathrm{max}(0,x)$. For the final fully connected layer no activation function was used. Since we are interested in binary classification we used the \texttt{logitbinarycrossentropy} in Flux, which is given by:
\begin{equation}\label{eq:logitbinarycrossentropy}
    H(q) = -\sum_{i=1}^N y_i\log(\sigma(q)) + (1-y_i)\log(1-\sigma(q)),
\end{equation}
where $\sigma(x) = (1+e^{-x})^{-1}$ and $y_i$ are the labels (1 for an image that waffled and 0 otherwise). This choice of the loss function is equivalent to using a sigmoid activation function in the last layer of the neural network but has better numerical stability.

CNNs are a form of supervised learning. Therefore, we first had to label a subset of the image reconstructions by hand. To find the labels, $y_i$, we analyzed 5000 random images from the scaled GRMHD set and an additional 500 from the stretched set. We then classified each image by whether it visually had a dominant ring-like feature or not. Some examples of images that passed and failed are shown in \autoref{fig:waffler}. Two-thirds of these classified images were used for training, and the rest for our test set. To combat overfitting, we also augmented the images by adding Gaussian random noise to each image when evaluating the loss function. Finally, ADAM \citep{ADAM} was used to optimize the network using the options defined in the Flux model zoo package\footnote{\url{https://github.com/FluxML/model-zoo/blob/master/vision/mnist/conv.jl}} with some minor changes. Namely, we broke our images into batches, with each batch containing 256 images, and used a learning rate of $3\times 10^{-3}$. However, if predictive performance on the training set did not improve after 10 epochs we dropped the learning rate by a factor of 10. The optimizer was run for 100 epochs and achieved an accuracy of 94\% and 92\% on the training and testing sets, respectively.

\end{document}